# Demystifying the 7-D Convolution Loop Nest for Data and Instruction Streaming in Reconfigurable AI Accelerators

Md Rownak Hossain Chowdhury, Mostafizur Rahman
*Division of Energy, Matters and Systems, Uuniversity of Missouri-Kansas City (UMKC)*
Kansas City, MO, US
{rhctmc, rahmanmo} @umkc.edu

*Abstract*— Convolution remains the most compute-intensive operation in AI acceleration, often constituting over 80-90% of the workload. Existing approaches in spatial architectures such as coarse-grained reconfigurable arrays (CGRAs) and field-programmable gate arrays (FPGAs) frequently rely on loop unrolling or GEMM-based matrix transformations, introducing significant overhead in both data movement and instruction control. This paper presents a new framework designed to systematically demystify the 7-dimensional convolution loop nest by reinterpreting it as a hardware-centric data and instruction streaming problem. Instead of treating the loop nest as a fixed computational construct, our approach exposes its structure as a set of spatial and temporal mappings governed by hardware parameters such as compute element distribution, interconnect topology, and reconfigurability. This abstraction supports lightweight, flexible deployment of convolution without reliance on heavyweight transformations or reordering schemes. We demonstrate the application of our approach on the MAVeC accelerator. We detail the implementation of convolution operations in MAVeC and extend the framework to support full model execution on VGG-16. Our profiling reveals high PE utilization (over 90%), significant fold reuse, and scalable throughput up to 1.56 TFLOPs/sec and 12.7 KIPS for end-to-end VGG-16 inference. These results validate the efficacy of our approach in minimizing control overhead, improving data locality, and enabling efficient large-scale convolution execution without reliance on conventional transformation-based methods.

## I. INTRODUCTION

Deep convolutional neural networks (CNNs) impose significant computational demands due to their inherently multidimensional computations [1]. These workloads place stress not only on arithmetic pipelines but also on data movement and instruction scheduling across hardware fabrics [2]. On general-purpose architectures such as central processing units (CPUs) and graphics processing units (GPUs), compilers attempt to optimize convolutional operations using techniques like loop unrolling, kernel fusion, and vectorization [3]. While these transformations improve instruction-level parallelism, they often flatten or restructure the computation in ways that conceal the original loop semantics, reducing visibility into data reuse patterns and limiting the ability to optimize spatial locality or control data routing [4].

A common consequence of such restructuring is the conversion of convolutions into two-dimensional matrix multiplications using general matrix-matrix multiplication (GEMM) [5]. This flattening enables reuse of optimized linear algebra libraries but discards the structure of the original seven-dimensional loop nest, which spans batch size, channel depth, kernel dimensions, and spatial resolution [6]. As a result, the computation loses natural opportunities for reuse and predictable scheduling, leading to increased off-chip memory traffic, irregular communication, and poor processing element (PE) utilization—particularly on spatial accelerators that depend on localized and structured dataflow for scalability [7].

Reconfigurable spatial accelerators such as FPGAs [8] and CGRAs [9] offer the architectural flexibility needed to exploit convolutional loop structures more effectively. By exposing fine-grained control over parallelism, memory access, and routing, these platforms allow designers to directly implement computation and communication patterns that preserve data reuse and locality [10]. While this flexibility is promising, current mapping flows for reconfigurable hardware often rely on kernel flattening, coarse tiling, or full bitstream reconfiguration to adapt to different layers [11]. Such approaches disrupt reuse opportunities, introduce reconfiguration delays, and require large configuration memories [12]. Additionally, accommodating variations in layer dimensions frequently involves complex host-side orchestration and can lead to inefficient resource utilization when layer shapes differ from the precompiled design [13].

To address these limitations, we introduce a principled seven-dimensional loop-nest decomposition and mapping strategy that preserves convolution semantics while harnessing the adaptability of reconfigurable hardware. The decomposition is expressed through three hardware-centric abstractions: filter folds, image folds, and image blocks. These abstractions define spatially structured units of computation that are mapped onto a two-dimensional array of processing elements. Execution follows a message-driven, weight-stationary model in which 64-bit messages carry both data and opcodes, enabling decentralized scheduling without global control. The architecture supports localized buffering to reduce memory access latency, spatial multicast to distribute weights efficiently, and in-network reduction to accumulate partial sums near their origin. These features minimize off-chip traffic and sustain high utilization across diverse convolutional layers. We implement and evaluate this design on MAVeC, a reconfigurable architecture that supports fine-grained streaming of both instructions and data [14], [15], [16]. Overall, the key contributions of this paper are as follows:

- Loop-nest decomposition that maps 7-D convolution space into 3-D abstractions, preserving reuse and parallelism boundaries.
- Data and instruction-streaming execution framework minimizing host intervention.

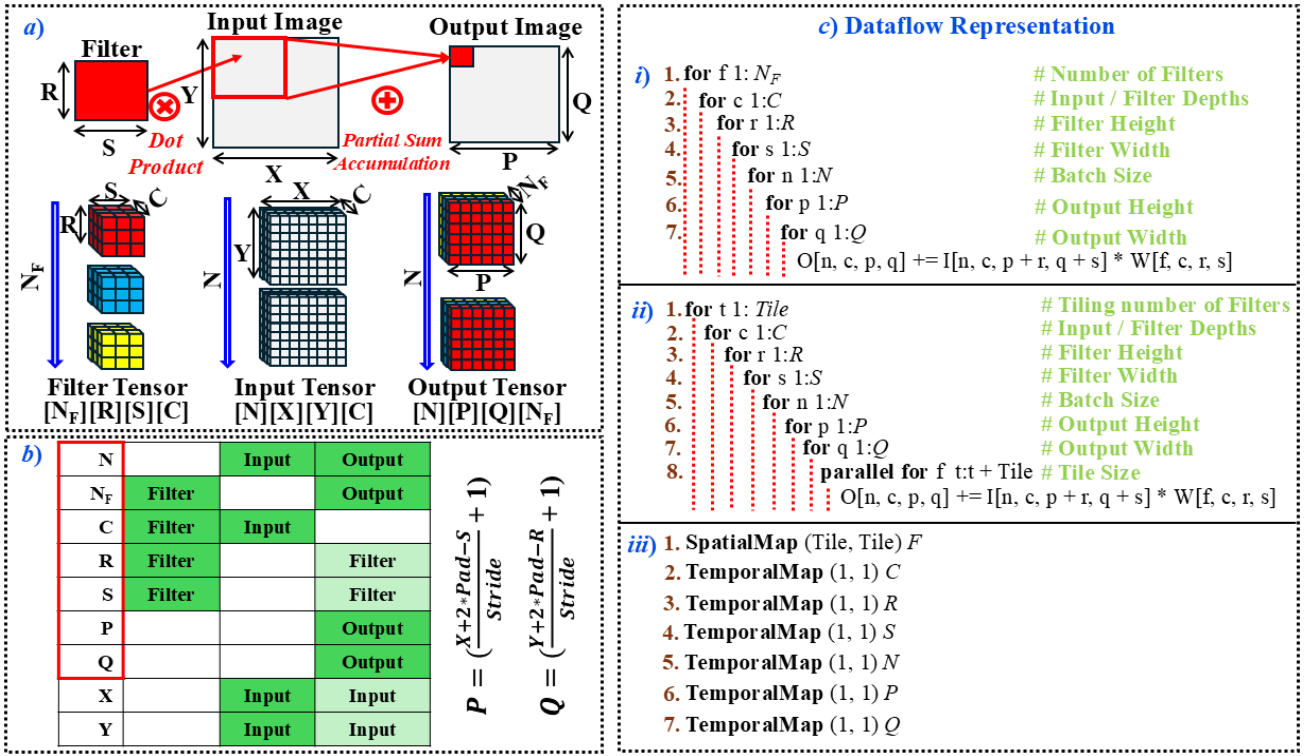

**Figure 1. (a) Convolution structure** illustrating spatial overlap, dot product, and accumulation across filter, input, and output tensors. **(b) Parameter alignment** across tensors, showing shared and derived dimensions in convolution. **(c) Dataflow representations:** (i) canonical 7D loop nest, (ii) tiled and parallelized loop nest, and (iii) data-centric directives illustrating a weight-stationary mapping strategy.

- Fine-grained on-chip routing scheme to support spatial multicast, local reduction, data reuse, and dynamic message hopping.
- An analytical performance model that unifies operational and communication cycles and predicts PE utilization, latency, and system throughput (KIPS).
- Empirical validation on VGG-16 and scaled workloads, achieving 12.7 KIPS and >90% utilization on a 64×64 array with GDDR7 and PCIe Gen6 at 1 GHz.

This work is organized as follows. Section *II* reviews related work on dataflow accelerators and reconfigurable hardware. Section *III* formalizes the 7D-convolution loop nest and identifies mapping opportunities for spatial dataflow architecture. Section *IV* details our decomposition strategy and its integration with MAVeC's message-driven fabric. Section *V* presents benchmarking results on VGG-16 and scaled workloads. Finally, section *VI* concludes with key findings and future directions.

## II. RELEVANT WORK IN LITERATURE

A dominant trend in CNN execution is the transformation of convolutions into General Matrix-Matrix Multiplication (GEMM) workloads using techniques such as im2col [17], Winograd [18], or FFT-based algorithms [19]. These transformations are widely adopted in CPU/GPU libraries like cuDNN [20] and MKL-DNN [21] due to their compatibility with optimized matrix-multiplication routines. However, these methods collapse spatial and channel dimensions into flat matrices, disrupting the inherent reuse structure of the original loops and making execution behavior opaque—especially in custom or spatial accelerators [22].

Even custom ASIC dataflow accelerators like TPU execute convolutions by lowering the input tensor into a 2D matrix form (e.g., via im2col or Toeplitz encoding) and stream it into a systolic array alongside filter weights [23]. Although systolic arrays excel at regular, matrix-style reuse, the transformation introduces data duplication, alignment overheads, and loss of semantic correspondence to the original loop structure [24]. Reuse boundaries become hard-coded, and fine-grained scheduling is offloaded to static compilation, limiting adaptability across diverse CNN layers [25].

In contrast to fixed-function accelerators, reconfigurable fabrics such as FPGAs and CGRAs offer more flexibility in execution. FPGA-based CNN accelerators rely on layer-specific bit-streams with static memory partitioning, requiring recompilation for any tensor shape change and often under-utilizing arithmetic resources [26]. CGRAs improve reconfiguration latency, but their modulo-scheduling over time-expanded graphs struggles with irregular strides, padding, or control, leading to inefficient pipeline utilization [27]. Moreover, layer transitions often require flushing data off-chip and reloading kernels, undermining temporal reuse and continuity across operations [28].

Other efforts, such as loop-nest-aware accelerators [29], [30], aim to exploit specific reuse patterns—row-stationary, weight-stationary, or output-stationary—by aligning loop axes with the physical layout of compute and memory. However, they depend on statically compiled schedules and centralized dispatch, making them rigid and less capable of supporting dynamic execution or cross-layer reuse [31], [32]. Similarly, compiler-driven frameworks like

TVM [*33*], Tiramisu [*34*], and Halide [*35*] explore loop transformations and autotuning, but typically reduce convolutions into lower-dimensional GEMMs. This abstraction hides execution order and reuse semantics, making them unsuitable for spatial hardware where explicit dataflow and control are essential [*36*].

In summary, existing hardware and compiler approaches fall into two broad categories: (i) matrix-based execution, which prioritizes throughput but discards structural reuse, and (ii) dataflow-specific mappings, which preserve some reuse but impose rigid, statically compiled flows. Neither class supports dynamic data and instruction streaming execution framework with embedded control. These limitations motivate our approach: a principled 7D decomposition that expresses each operation as a routed message—embedding control, reuse, and timing within the data itself—and enabling distributed scheduling over a spatial architecture minimizing centralized orchestration.

## III. 7-D CONVOLUTION LOOP NEST & OPPORTUNITIES FOR SPATIAL ARCHITECTURE BASED ACCELERATION

**a) Demystifying the 7D Loop Nest:** Convolution operations, particularly 3D convolutions used in deep learning models, operate over three primary tensors: the filter tensor ($N_F, R, S, C$), the input tensor ($N, X, Y, C$), and the output tensor ($N, P, Q, N_F$), as shown in *Figure 1(a)*. Here, $N_F$ denotes the number of filters, ($R, S$) defines the spatial extent of each filter, $C$ is the input channel count, ($X, Y$) represents the input resolution, and $N$ is the batch size. During convolution operation, each filter slides across the spatial dimensions of the input tensor, performing element-wise multiplications between the filter and corresponding input window, and the results are accumulated to produce partial sum, as depicted in *Figure 1(a)*.

Sliding the filters across the spatial positions of the input tensor introduces structural overlap and dimensional dependencies among the participating tensors. These relationships are summarized in *Figure 1(b)*, where each row corresponds to a tensor variable, and each column indicates its presence in the filter, input, or output tensor. For instance, the batch size $N$ appears in both input and output tensors, $N_F$ is shared between filters and output, and $C$ is common to both filters and inputs. Moreover, the output dimensions $P$ and $Q$ are not independent; rather, they are derived from the input size ($X, Y$), filter size ($R, S$), stride, and padding configuration. Collectively, these relationships define the complete indexing structure of the convolution operation, which can be fully expressed using seven parameters: $N, N_F, C, R, S, P$, and $Q$. Each of these parameters maps directly to a loop in a 7-level nested iteration space, with each loop iterating over one of these dimensions [*37*]. This formulation captures the entire computation space of a convolution layer and provides a unified view for analysis and mapping.

**b) Acceleration Opportunities in dataflow Architecture:** Spatial dataflow hardware architectures are purpose-built to accelerate convolution workloads by exploiting both parallelism and data reuse. A typical spatial accelerator consists of a two-dimensional ($R_P$ X $C_P$) array of processing elements (PEs). Each PE includes a floating-point unit (FPU), and a small local buffer used to store filter weights, input activations, or partial sums during computation [*38*]. The number of PEs determines the degree of spatial parallelism, while buffer size and numerical precision (e.g., FP32 or INT8) affect the efficiency of data retention and computation. This PE array operates under a broader memory hierarchy. On-chip memory levels serve as staging areas for moving data between the compute fabric and off-chip memory, thereby reducing bandwidth demands [*39*]. A network-on-chip (NoC) typically implemented using buses, trees, or store-and-forward topologies, enables data movement both among PEs and between memory levels [*40*]. This architectural arrangement supports both spatial and temporal data reuse, which is critical for efficient execution of convolution operations.

The execution of convolution on such architectures is governed by dataflow [*41*], which defines the spatiotemporal schedule of operations and data movement. It determines which data is prioritized in the memory hierarchy, how often values are reused, and which forms of parallelism are activated in the compute fabric. There are three widely used representations of dataflow that help express these mappings. The first is the canonical 7D loop nest [*42*], shown in *Figure 1(c-i)*, where loops iterate over filters (**F**), channels (**C**), filter dimensions (**R, S**), batch size (**N**), and output spatial dimensions (**P, Q**). The loop order affects data reuse; for example, filter-first ordering enables weight stationarity, where weights remain fixed in PEs while other data streams through [*43*].

To make parallelism and data reuse more explicit, the second representation introduces loop tiling and parallel-for constructs [*44*], illustrated in *Figure 1(c-ii)*. Here, each dimension is partitioned into tiles to fit within local memory or PEs. Outer loops enumerate tiles, while inner loops compute over tile elements. The use of "**parallel for**" indicates which tiles are executed concurrently across PEs, while standard "**for**" loops imply temporal iteration. This model reflects how the computation is distributed across hardware resources and managed by the memory hierarchy. The third representation is a data-centric directive model [*45*], shown in *Figure 1(c-iii)*, which focuses on how data—not control—is assigned and moved across PEs. For instance, a directive like "**Spatial Map (Tile, Tile) F**" distributes filter indices across the PE array, while directives such as "**Temporal Map (1, 1) C**" or "**Temporal Map (1, 1) P**" specify serialized processing within each PE. This abstraction allows designers to describe data movement and reuse patterns explicitly, without embedding those decisions into control logic or rigid loop orderings.

By aligning the seven-dimensional convolution loop space with the architectural models, designers can uncover substantial opportunities to reduce memory access overhead, improve throughput, and fully utilize hardware parallelism. This understanding forms the operational core of our framework, which leverages the convolution's inherent structure to orchestrate efficient data and instruction streaming across spatial accelerators.

## IV. OUR APPROACH FOR PARRALELLIZING CONVOLTUION & MAPPING ONTO MAVeC'S SPATIAL ARCHITECTURE

### A. MAVeC Fabric:

The Messaging-based Adaptive Vector Computing (MAVeC) fabric, illustrated in *Figure 2*, operates within a

System-on-Chip (SoC) environment that unifies computing, memory, and communication subsystems. A central system bus links the host processor, off-chip memory, and MAVeC through a PCIe interface. Data and instructions are first stored in off-chip memory and transferred into MAVeC's three-tier on-chip memory system—comprising Tile Buffers, SiteM Memories, and SiteO local storage—designed for low-latency access and progressive data staging. At the compute layer, each core consists of four quads, each with 256 Tiles arranged in a 16X16 mesh. Each Tile includes 16 SiteMs, and each SiteM houses a 4X4 grid of SiteOs, the fundamental processing element (PE) equipped with FPUs, dual input FIFOs, and local buffers.

MAVeC follows a message-driven execution model, where 64-bit messages encode operation and routing metadata, enabling pipelined compute and data movement. Messages are routed across a 2D mesh via destination-aware logic, supporting streaming and forwarding behaviors. To facilitate high-throughput dataflow, SiteMs implement spatial buses for multicasting and reduction, recursively extended across hierarchy levels. This architectural foundation was detailed in our prior works [14], [15], [16], where we introduced the MAVeC microarchitecture, defined its instruction set architecture (ISA) and message encoding format, and described its hierarchical memory subsystem. We further benchmarked its performance on convolution workloads under TSMC 28 nm technology node. Building on that foundation, this work aims to elaborate the dataflow and mapping strategy for convolutional workloads.

### B. 7D Loop Nest Decomposition on MAVeC

We reinterpret the canonical seven-dimensional (7D) loop nest into a hardware-conscious execution model to map convolutional workloads onto MAVeC's spatial architecture. This decomposition aligns algorithmic structure with MAVeC's architectural constraints, unlocking fine-grained parallelism, enhancing data reuse, and minimizing off-chip memory access—all while adhering to its message-driven execution paradigm. Central to this mapping strategy is the reorganization of the 7D iteration space into three core constructs: Filter Folds (FF), Image Folds (IF), and Image Blocks (IB). These abstractions decouple stationary and streamed components—keeping filter weights fixed on the PE (SiteO) array while image data is dynamically injected in tiled, depth-aware segments. Their synchronized interaction yields partial sums, which are accumulated across the depth dimension to compute the final output tensor. The remainder of this section elaborates each stage of this decomposition pipeline—from fold generation and block partitioning to spatial execution and output reduction.

***a) Filter Fold and Image Block Generation:*** The transformation from a high-dimensional filter tensor to a format compatible with MAVeC's PE array begins by flattening the 4D filter tensor into a 2D matrix to enable parallel execution and spatial reduction. The flattening process adopts a depth-major traversal, where each filter channel $C$ is processed sequentially. Within each channel, the spatial grid ($RXS$) is unrolled by traversing columns in reverse order—from the last to the first—capturing elements top to bottom within each column. To facilitate column-wise reduction across rows during execution, an additional column is inserted after each spatial row ($R$), expanding the effective width from $S$ to $S+1$. All filters are then stacked row-wise, forming a 2D matrix of shape ($N_F$, $C_{Transformed}$), where $C_{Transformed} = C \times R \times (S+1)$.

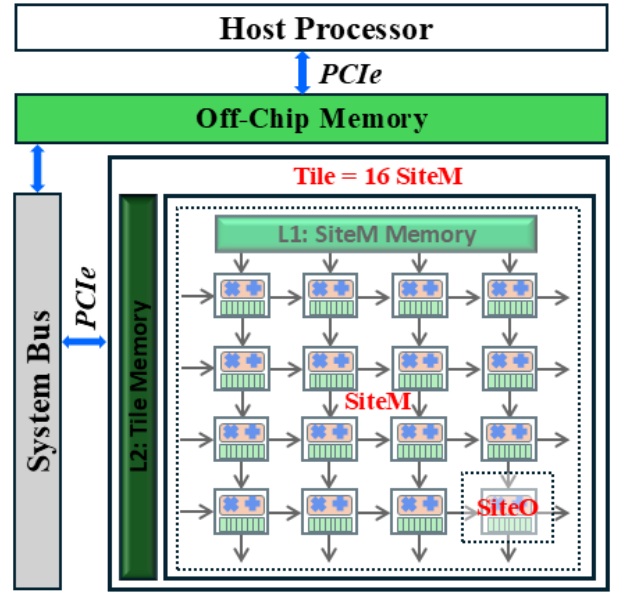

**Figure 2. System-on-Chip integration of MAVeC.** PCIe connects the host processor to MAVeC and transfers data to its three-level on-chip memory hierarchy: Tile Buffers (L2), SiteM Memories (L1), and L0 local storage in SiteOs. Each Tile contains 16 SiteMs, and each SiteM includes a 4×4 grid of SiteOs with FPUs and local buffers.

This transformation is demonstrated in the top and middle portion of *Figure 3(a)*, which depicts a filter tensor comprising $N_F = 4$ filters, each with $C = 4$ channels and a spatial extent of $RXS = 3X3$. During flattening, each channel is traversed in depth-major order ($C = 0$ to $3$), and its spatial region is unrolled column-by-column in reverse—starting from the last spatial column ($S = 2$) and proceeding leftward to the first ($S = 0$). Within each column, elements are read top-to-bottom across the rows ($R = 0, 1, 2$). For example, in the first row of the transformed matrix, the rightmost colored segment represents channel $C = 0$ of filter 1 (red). The initial group of entries $\{F_2^{(0)}, F_5^{(0)}, F_8^{(0)}\}$ originates from the 3$^{rd}$ column ($S = 2$), followed by a reserved column (gray). This is followed by $\{F_1^{(0)}, F_4^{(0)}, F_7^{(0)}\}$ from column $S = 1$, again followed by padding, and finally $\{F_0^{(0)}, F_3^{(0)}, F_6^{(0)}\}$ from column $S = 0$, with another reserved column. This column-reversed unrolling and interleaved padding pattern is applied uniformly to the remaining channels ($C = 1$ to $C = 3$). Thus, each filter occupies $C_{Transformed} = C \times R \times (S+1) = 4 \times 3 \times (3+1) = 48$ columns after flattening. The same transformation is applied to the other three filters, producing a total of $R_{Transformed} = N_F = 4$ rows in the transformed matrix.

Following the transformation of the filter tensor into a 2D matrix, the next step involves slicing this matrix into filter folds compatible with MAVeC's PE array, as depicted in the bottom portion of *Figure 3(a)*. This slicing is governed by the architectural dimensions of the PE array: the fold height is set to the number of rows ($R_P$) in the PE array, as defined in **equation (1)**. The fold width is calculated using **equation (2)**, which determines the number of depth slices that can fit horizontally within the available PE columns ($C_P$). This is calculated by first determining how many $R * (S + 1)$-sized depth slices fit within $C_P$, and then scaling by the slice width. Consequently, the total number of filter folds is given by **equation (3)**, which multiplies the number of vertical splits (along filters) with the number of horizontal splits (along depth slices).

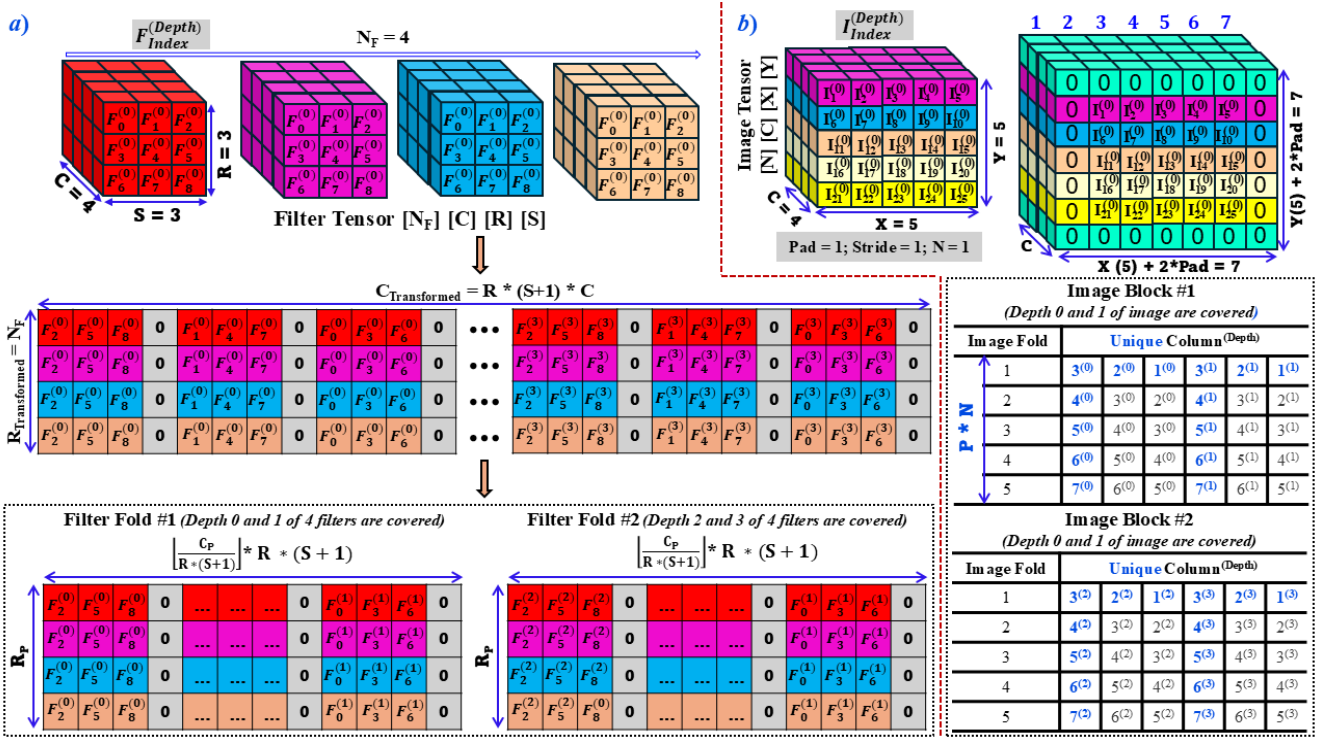

**Figure 3. Filter and image fold generation under hardware constraints. (a)** The 4D filter tensor is flattened into a 2D matrix with additional columns reserved for spatial reduction. This matrix is sliced into filter folds based on available PE array dimensions. **(b)** The image tensor is folded along depth and width to mirror the structure of filter folds. Image folds sharing the same depth range are grouped into blocks, and any columns reused across folds are omitted in subsequent folds.

The example illustrated in *Figure 3(a)* attempts to map the transformed matrix ($R_{Transformed} = 4$, $C_{Transformed} = 48$) onto a PE array with dimensions $R_P = 4$, $C_P = 24$. Here, each depth slice (i.e. 1 filter channel) spans $R * (S + 1) = 3 * (3 + 1) = 12$ columns, allowing $\lfloor C_P/R.(S+1) \rfloor = \lfloor 24/12 \rfloor = 2$ channels per fold, yielding a fold width of $2 \times 12 = 24$ columns. Since the PE array can cover all four filters (as $R_P = N_F = 4$), only one horizontal split is required. However, the 48 columns of the transformed matrix necessitate two vertical splits. As a result, two filter folds are produced, each being a $4 \times 24$ submatrix: Filter Fold #1 spans channels $C = 0, 1$ (columns 0–23), and Filter Fold #2 spans $C = 2, 3$ (columns 24–47).

Following the filter folding process, the 4D image tensor ($N, C, X, Y$) undergoes depth-wise ($C$) and width-wise ($X$) partitioning. First, depth-wise slicing divides the tensor into image blocks, where each block spans a group of channels equal to the depth coverage of a filter fold ($\lfloor C_P/R.(S+1) \rfloor$). Hence, the total number of image blocks equals the total number of filter folds, as expressed in **equation (4)**. Next, within each image block, the image is further partitioned along the width dimension to generate image folds. The number of folds per block for a single image corresponds to the convolution output width ($P$), and the total number of image folds per block across all $N$ images is $P \times N$,

$$\text{Filter Fold}_{Rows} = R_P \quad (1)$$

$$\text{Filter Fold}_{Columns} = \left\lfloor \frac{C_P}{R.(S+1)} \right\rfloor * R * (S+1) \quad (2)$$

$$\text{Total}_{Filter\ Folds} = \left\lfloor \frac{N_F}{\text{Filter Fold}_{Rows}} \right\rfloor * \left\lfloor \frac{C_{Transformed}}{\text{Filter Fold}_{Columns}} \right\rfloor \quad (3)$$

$$\text{Total}_{Image\ Blocks} = \text{Total}_{Filter\ Folds} \quad (4)$$

$$\text{Image Fold}_{Per\ Image\ Block} = P * N \quad (5)$$

as defined by **equation (5)**. Each image fold captures a spatial receptive field of width $S$ across all its depth slices. For the $i^{th}$ fold, the candidate columns selected are defined as $\{C_i, C_i + 1, ..., C_i + S - 1\}$, where $C_i = i * stride$ and $i \in \{1, 2, ..., P\}$. These columns are then reversed to align with the filter fold layout. If any column has already been used in a previous fold, it is excluded, ensuring only unique columns are retained per fold to minimize redundant data movement.

In *Figure 3(b)*, the image tensor has shape ($N = 1$, $C = 4$, $X = 5$, $Y = 5$), and padding is applied to yield spatial dimensions $7 \times 7$. Here, each image block accommodates $\lfloor C_P/R.(S+1) \rfloor = \lfloor 24/12 \rfloor = 2$ channels, leading to two image blocks: Block#1 covers channels 0 to 1, and Block#2 covers channels 2 to 3. Since the convolution output width ($P$) is 5, and $N = 1$, each block contains $P \times N = 5$ image folds. Each image fold selects $S = 3$ spatial columns per depth. For example, Fold #1 of Block #1 selects columns $\{3, 2, 1\}$ from both depth 0 and 1; Fold #2 starts at column 2, yielding $\{2, 3, 4\}$, but retains only the unique column $\{4\}$. This continues with Folds #3 to #5 selecting new columns $\{5\}$, $\{6\}$, and $\{7\}$, respectively. Block #2 follows the same logic for depths 2 and 3.

*b) Filter Fold and Image Block Interaction:* Once filter folds and image blocks are generated, the subsequent stage

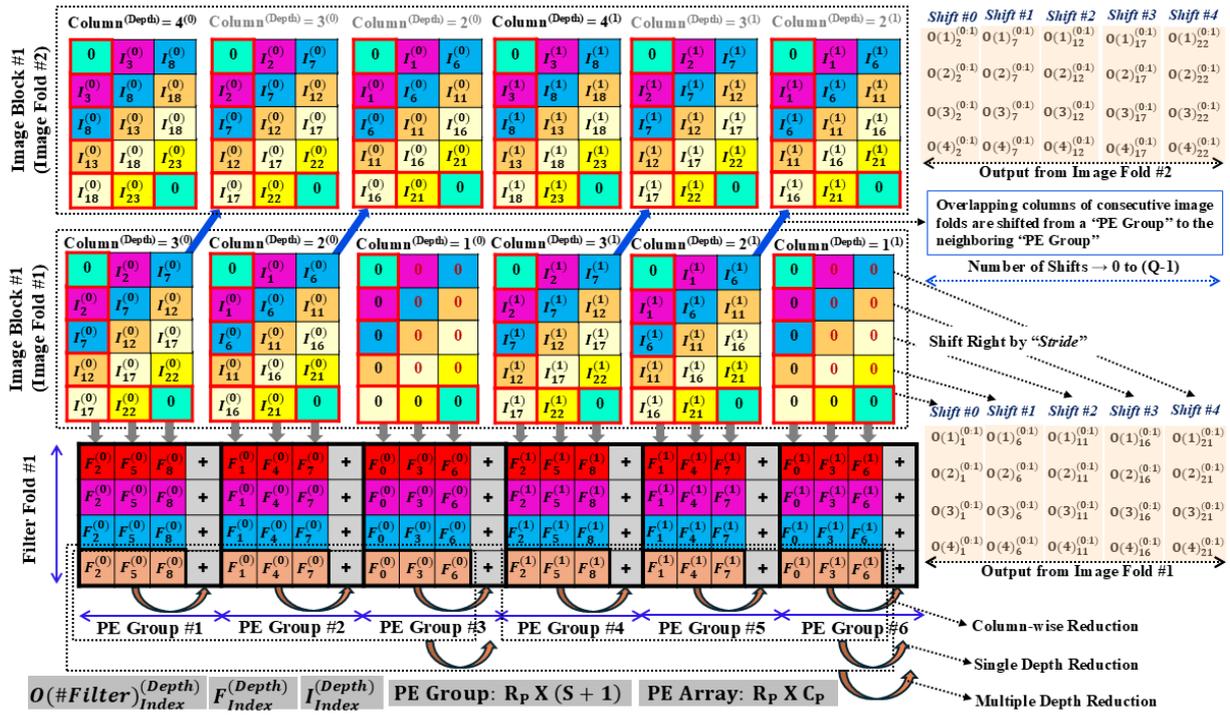

**Figure 4. Filter and image interaction in the PE array.** The filter fold is first spatially mapped onto the PE array and remains stationary throughout all image folds in a block, enabling weight reuse. Each image fold is then sequentially (1) multicast across rows, (2) element-wise multiplied with resident weights, (3) spatially reduced across the filter width and depth, and (4) right-shifted by the stride to generate the next output index. Over Q shifts, each fold produces a complete output column, with overlapping columns propagated laterally across PE groups.

involves their synchronized interaction to compute the output. Each image fold within an image block is sequentially streamed to the PE array, while the corresponding filter fold remains stationary throughout the interaction. This process unfolds in five stages: (1) programming the PE array with a stationary filter fold (2) spatially multicasting an image fold across all hardware rows (3) performing elementwise multiplication across the PE array (4) executing spatial reduction on the resulting partial sums and (5) right-shifting the image fold by convolution stride across PE groups.

The execution pipeline begins by configuring the PE array with a single filter fold, which remains fixed throughout its interaction with an image block. As illustrated in the bottom portion of *Figure 4*, filter fold #1, with the shape of 4 X 24, is spatially mapped onto a PE array. Here, each row of the array corresponds to a distinct filter, while each column holds to a spatially flattened weight element derived from the filter tensor. This spatial mapping enables fully parallel elementwise multiplication across the PE array during each interaction cycle with the streamed image fold. Moreover, keeping the filter weights stationary facilitates temporal reuse, as the same set of weights is repeatedly utilized across multiple input folds and shift cycles.

Once programmed, each column of the image fold is spatially multicast across PE groups, where each PE group spans $S+1$ columns. As illustrated in the bottom portion of *Figure 4*, the PE array is partitioned into six PE groups, each having 3 active columns and 1 reserved column. During each image fold interaction, designated columns from the image fold are assigned to these PE group. As shown in the middle portion of *Figure 4*, the first, second, and third PE groups receive columns 3, 2, and 1 from depth 0, respectively. Simultaneously, the fourth, fifth, and sixth PE groups receive columns 3, 2, and 1 from depth 1. Within each group, $S$ elements of a column are vertically multicast in reverse spatial order—from the last element to the first—across the PEs. This vertical reverse-order broadcast occurs in parallel with all PE groups. This spatial broadcasting ensures that the same input is simultaneously applied to multiple filters distributed across different hardware, enabling concurrent elementwise multiplication.

Upon completing the elementwise multiplication, partial sums are reduced hierarchically across the PE array in three distinct stages. These reduction stages are annotated at the bottom of *Figure 4* using curved arrows, each indicating a different scope of reduction. First, column-wise reduction aggregates values horizontally along the filter width $S$ across PEs within each row. Next, single-depth reduction accumulates these results across column groups corresponding to the same depth slice to generate a unified output for each depth. Finally, multi-depth reduction combines results across multiple depths to produce a single scalar output per spatial position and filter. As a result, at shift #0, the complete set of partial sums for output index1, covering both depth 0 and 1, is computed for all filters.

Each image fold then undergoes a series of rightward shifts across the PE array, governed by the convolution stride. In each shift cycle, the columns of the fold are advanced by the stride unit and reinjected into the PE groups, enabling spatial reuse of input activations across neighboring regions. The total number of shift cycles is $Q$, corresponding to the number of output columns produced per filter. Each shift contributes one output index per filter over the assigned depth range. Thus, after $Q$ shifts of a given fold, a complete output column is generated for all filters across the targeted depth slices. As illustrated on the right side of *Figure 4*, five shift cycles yield one output column per filter for depths 0 and 1. Moreover, when consecutive image folds contain overlapping columns, these reused columns are propagated laterally across PE groups, as denoted by the blue arrows pointing from one PE group to the next. This reuse strategy ensures that previously injected data can be forwarded instead of redundantly reloaded, thereby enhancing data locality and reducing bandwidth requirements.

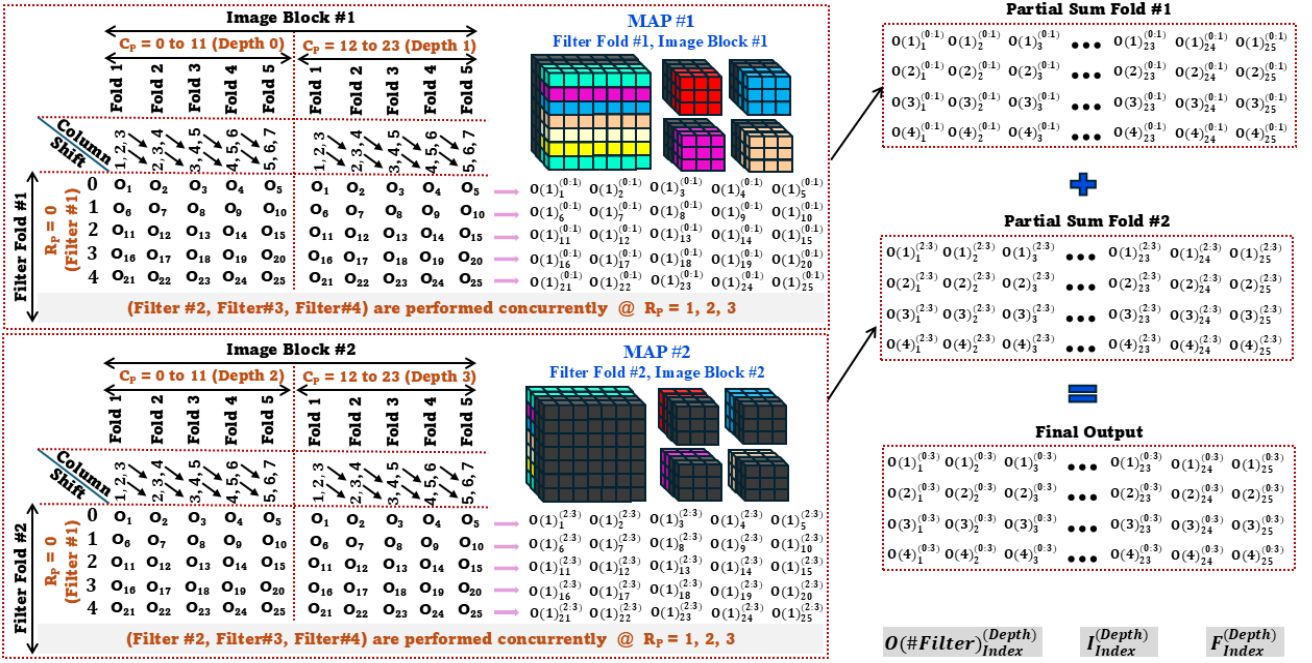

**Figure 5. Generating final output via partial sum accumulation.** Each MAP instance (e.g., MAP #1, MAP #2) corresponds to the interaction between a filter fold and an image block, producing partial sums over specific filter and depth ranges. As shown on the right, partial sum folds from different depth slices are aggregated to produce the final output tensor, which spans all filters and the complete input depth.

The reuse and parallelization metrics of this mapping strategy are formalized in **equations (6) to (9)**. First, temporal reuse (**Eq. 6**) of weights arises as each stationary filter fold is repeatedly applied across all shifts and folds within an image block, amortizing weight loading over ($P *Q$) cycles and yielding significant reuse across $R_P$ filters and $\lfloor C_P/R.(S+1) \rfloor * R * S$ columns of PE array. Second, spatial reuse of input activations (**Eq. 7**) is achieved by streaming each image fold once while enabling reuse across all $Q$ shift iterations and across $R_P * \lfloor C_P/R.(S+1) \rfloor * R * S$ PEs. Third, spatial parallelism (**Eq. 8**) quantifies the number of active PEs involved in each cycle of computation. Finally, during spatial reduction (**Eq. 9**) every output element, $S$ multiplication results per filter are first reduced along the filter width, followed by accumulation across $R$ PE groups (depth-wise) and then across depth slices (multi-depth) across all $P$ folds and $Q$ shifts per image block.

*c) Final Output Generation:* The convolution process yields intermediate partial outputs during the interaction between filter folds and image blocks. As depicted in *Figure 5*, each mapping instance (MAP) corresponds to a specific pairing of one filter fold with one image block. In MAP #1 (top of *Figure 5*), Filter Fold #1 comprises four filters mapped across PE rows $R_P = 0$ to $R_P = 3$. Each filter's first two depth slices—Depth 0 and Depth 1—are spatially placed along column partitions $C_P = 0\text{-}11$ and $C_P = 12\text{-}23$, respectively. Image Block #1 provides matching input values from Depth 0 and Depth 1, which are multicast to the corresponding $C_P$ partitions, enabling synchronized interaction with spatially mapped filters. MAP #2 (bottom of *Figure 5*) follows the same structure: Filter Fold #2, also spanning $R_P = 0\text{-}3$, interacts with Image Block #2, covering Depth 2 and Depth 3. The image elements from these depths are multicast to $C_P = 0\text{-}11$ and $C_P = 12\text{-}23$, respectively, aligning with the depth-sliced filter columns.

Each image block consists of five image folds ($P = 5$), and each fold undergoes five rightward shifts ($Q = 5$) governed by the convolution stride ($S = 1$). These folds are streamed sequentially into the PE array, where they interact with a stationary filter fold. During each shift of an image fold, a single output index is computed for all $R_P$ filters. Across all $Q$ shifts of that fold, a complete output column is produced for the same set of filters. Repeating this process over all $P$ image folds in a block yields all output columns, resulting in a total of **25 (P X Q)** output indices per filter per image block. The intermediate outputs from MAP #1 and MAP #2 are shown in the right panels of Fig. 5 as Partial Sum Fold #1 and Partial Sum Fold #2, respectively, representing

$$\text{Temporal}_{\text{Reuse (Weight)}} = P * Q * R_P * \left\lfloor \frac{C_P}{R*(S+1)} \right\rfloor * R * S \quad (6)$$

$$\text{Spatial}_{\text{Reuse (Input)}} = Q * R_P * \left\lfloor \frac{C_P}{R*(S+1)} \right\rfloor * R * S \quad (7)$$

$$\text{Spatial}_{\text{Parallelism (PE)}} = R_P * \left\lfloor \frac{C_P}{R*(S+1)} \right\rfloor * R * (S+1) \quad (8)$$

$$\text{Spatial}_{\text{Reduction (Partial Sum)}} = P * Q * R_P * \left\lfloor \frac{C_P}{R*(S+1)} \right\rfloor * S \quad (9)$$

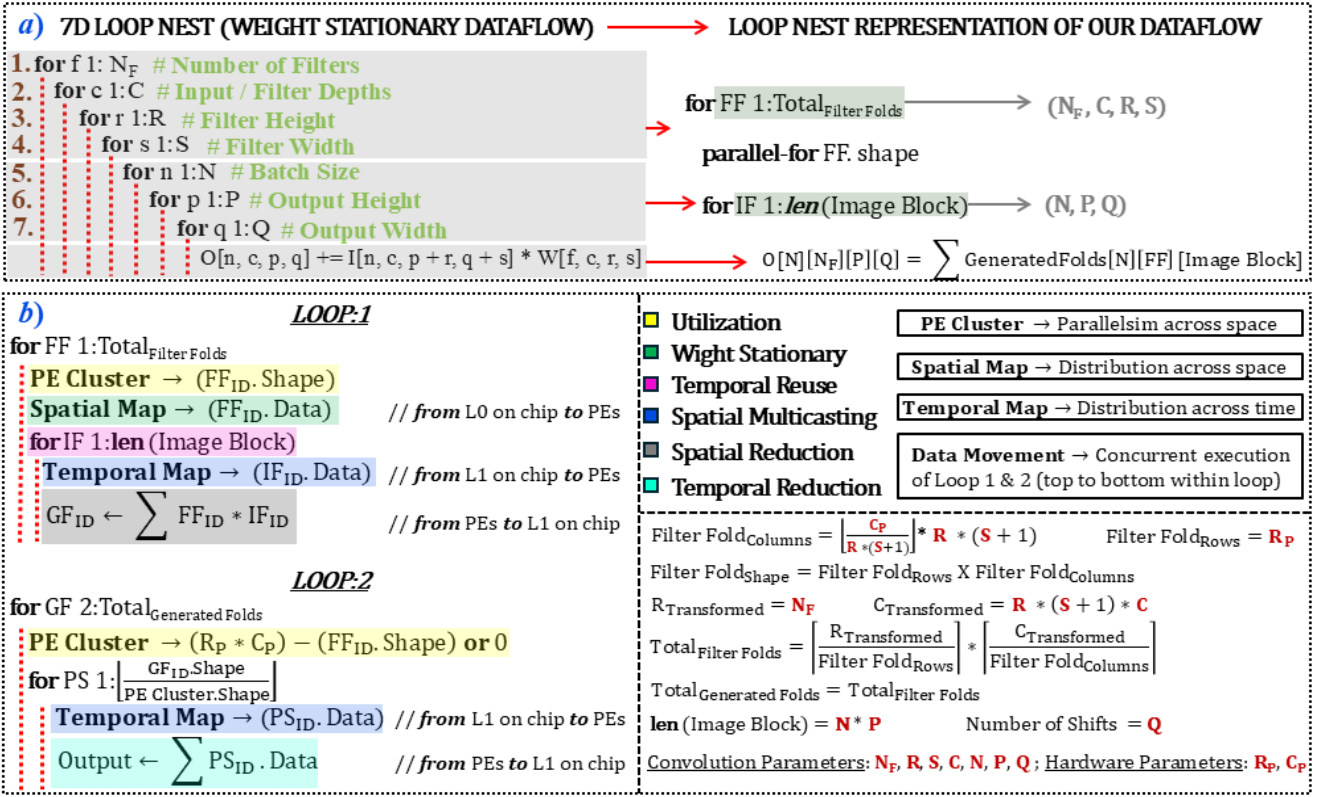

Figure 6. Proposed dataflow for 7D convolution workload using two complementary views. **(a)** The loop-nest formulation reinterprets the canonical 7D convolution space using three key constructs—Filter Fold (FF), Image Block (IB), and Image Fold (IF)—to expose spatial parallelism and temporal pipelining. **(b)** The data-centric view highlights the orchestration and reuse of data, with Loop 1 coordinating FF-IF interactions and Loop 2 reducing partial sums into final outputs.

contributions from depths 0–1 and 2–3. These partial results are stored in L1 memory and later accumulated along the depth dimension to form the final output tensor.

*d) Summary of Mapping Strategy:* **Figure 6(a)** demonstrates the transformation of the canonical seven-dimensional (7D) convolution loop nest—comprising filter ($N_F$), channel (C), kernel height and width (R, S), batch (N), and output spatial dimensions (P, Q)—into three primary mapping constructs: Filter Fold (FF), Image Block (IB), and Image Fold (IF). The filter fold merges the $N_F$, C, R, and S loops and is statically programmed onto the $R_P$ X $C_P$ processing array, facilitating weight-stationary execution through spatial tiling. The image block reorganizes the input dimensions N, X, and C into a grouped structure that spans both a depth-partitioned and width-sliced portion of the input, with an effective length of P X N. Within each block, the image fold divides this region into Q temporally shifted subsegments, where shift length is determined by the stride. Spatial parallelism is exposed via parallel-for constructs over FF, while temporal pipelining is realized through sequential for iterations across IB and within each IF shift stage.

**Figure 6(b)** complements loop-based abstraction with a data-centric view that explicitly captures data movement, reuse, and scheduling behavior across hardware levels. The mapping constructs—Filter Fold (FF), Image Fold (IF), and Partial Sum Fold (PS)—are orchestrated through two independent nested loops. In **Loop 1**, FFs are statically distributed across PE clusters via spatial mapping, while IFs are streamed temporally to interact with each FF, enabling reuse of weights across Q shifts and P image folds. Each IF is multicast along the $R_P$ rows, exploiting spatial reuse of activations. **Loop 2** accumulates the resulting PS folds, which are shaped by the interaction between FF and IF, through a sequence of depth-wise reductions. These partial outputs are temporally reduced to finalize the output tensor. The data-centric directives—**Spatial Map** and **Temporal Map**—explicitly encode the movement of data between on-chip memory and compute array.

**Table 1** summarizes the proposed dataflow by consolidating the key mapping constructs, their associated parameter spaces, execution strategies, and reuse mechanisms. Each construct—from the spatially tiled PE cluster to the temporally shifted image folds—plays a distinct role in aligning the original 7D loop nest with hardware execution. Filter folds exploit weight reuse across shifts and blocks, while image folds enable input reuse via multicasting and stride-based traversal. Image blocks organize depth-wise streaming, and partial sums are locally reduced before final accumulation.

## V. BENCHMARKING

### A. Experimental Setup

We evaluate the proposed mapping strategy on the MAVeC AI accelerator, a spatially programmable ASIC architecture developed under the TSMC 28 nm technology node. All experiments are carried out using a custom simulator that accurately models MAVeC's micro-architectural behavior, including compute data-path, hierarchical instruction streaming, and message-driven execution. Folding and scheduling primitives embedded in the simulator reflect the semantics of the proposed mapping strategy. For this study, the accelerator is instantiated in three spatial configurations—16 × 16, 32 × 32, and 64 × 64 processing elements (PEs)—to quantify scalability with increasing parallelism. Each PE executes single-precision (FP32) operations at 1 GHz, while system I/O is provisioned by DDR7 off-chip memory and a PCIe x16 Gen host interface. We assess the performance using three distinct metrics: (i) average PE utilization (%), (ii) total execution time (KCCs), and (iii) throughput (GFLOPs/second). These metrics are gathered for two workload classes: (a) a synthetic 3 × 3 convolution suite with varying channel dimensions (*Table 2 (A)*), and (b) a layer-wise profile of VGG-16 with batch size 1 (*Table 2 (B)*). For system-level assessment, we present the end-to-end throughput of VGG-16 in Kilo Inferences Per Second (KIPS).

Table 1. Summary of Mapping Constructs, Execution Strategy, and Reuse Behavior

| Mapping Constructs | Parametric Space | Mapping Strategy | Reuse/Movement |
|---|---|---|---|
| PE Cluster | $R_P$, $C_P$ | Tiling for parallel execution | Spatial mapping |
| Filter Fold (FF) | $N_F$, C, R, S, $R_P$, $C_P$ | Stationary on PE array | Temporal reuse |
| | | $N_F$ mapped to $R_P$ | |
| | | C, R, S mapped across $C_P$ | |
| Image Fold (IF) | C, Pad, R, S, Q, Stride | Depth (C) aligned with filter fold | Spatial reuse (Multicasting) Temporal Reuse (Shifts) |
| | | Extract R columns from a depth | |
| | | Sliced by S | |
| | | Shift right by stride | |
| | | Q shifts per fold | |
| | | Multicast across $R_P$ | |
| Image Block | N, C, P | Fold grouped by depth (C) | Temporal mapping |
| | | Block length is P * N | |
| | | Schedule each fold sequentially | |
| Partial Sum Fold | Derived from {FF, IF} | Accumulation across depth (C) | Spatial reduction |
| Final Accumulation | $N_F$, P, Q, N | Sum partial sum folds | Temporal reduction |

**Table 2. Convolutional workloads used for evaluation**
A) Synthetic 3×3 convolution suite with varying depth and filter count. B) Layer-wise breakdown of VGG-16 convolution layers.

| No. | Image Tensor (I X I X D) | Filter Tensor (F X F X D X $N_F$) | | |
|---|---|---|---|---|
| A) | | | | |
| 1 | 56 X 56 X 64 | 3 X 3 X 64 X 64 | Stride = 1 | Pad = 1 |
| 2 | 56 X 56 X 128 | 3 X 3 X 128 X 128 | | |
| 3 | 56 X 56 X 256 | 3 X 3 X 256 X 256 | | |
| 4 | 56 X 56 X 512 | 3 X 3 X 512 X 512 | | |
| B) | | | | |
| 1.1 | 224 X 224 X 3 | 3 X 3 X 3 X 64 | Stride = 1 | Pad = 1 |
| 1.2 | 224 X 224 X 64 | 3 X 3 X 64 X 64 | | |
| 2.1 | 112 X 112 X 64 | 3 X 3 X 64 X 128 | | |
| 2.2 | 112 X 112 X 128 | 3 X 3 X 128 X 128 | | |
| 3.1 | 56 X 56 X 128 | 3 X 3 X 128 X 256 | | |
| 3.2 | 56 X 56 X 256 | 3 X 3 X 256 X 256 | | |
| 3.3 | 56 X 56 X 256 | 3 X 3 X 256 X 256 | | |
| 4.1 | 28 X 28 X 256 | 3 X 3 X 256 X 512 | | |
| 4.2 | 28 X 28 X 512 | 3 X 3 X 512 X 512 | | |
| 4.3 | 28 X 28 X 512 | 3 X 3 X 512 X 512 | | |
| 5.1 | 14 X 14 X 512 | 3 X 3 X 512 X 512 | | |
| 5.2 | 14 X 14 X 512 | 3 X 3 X 512 X 512 | | |
| 5.3 | 14 X 14 X 512 | 3 X 3 X 512 X 512 | | |

*B. Evaluation Metrics*

To analyze the proposed mapping strategy, we formulate a set of analytical expressions that incorporate mapping constructs such as fold dimensions, fold counts, and routing intervals to estimate performance metrics such as utilization, execution latency, and system throughput. **Equation (10)** defines the average utilization, which quantifies how efficiently processing elements (PEs) are spatially packed across filter folds. For each fold, the number of active PEs is normalized by the full array size ($R_P$ X $C_P$), and the average is taken across all folds ($N_{FT}$). Unutilized sites ($Idle_i$) are dynamically calculated based on each filter fold shape.

$$\text{Util}_{avg}(\%) = \frac{1}{N_{FT}} \sum_{i=1}^{N_{FT}} \frac{(R_P \text{ X } C_P) - (\text{Idle})_i}{(R_P \text{ X } C_P)} \quad (10)$$

**Equation (11)** models the total execution time ($T_{Ops}$) capturing both compute and communication overheads. Here, $N_{FT(C)}$ represents the number of filter folds along the column dimension, while $N_{FT(R)}$ captures the row-wise tiling. The term ($4 \cdot \text{Shifts} \cdot N_{DT} \cdot N_{FT(C)}$) accounts for horizontal shifts of all image folds ($N_{DT}$), and $K = \log_{(I+1)}(C_P) + 1$) estimates routing latency through reserved columns. Finally, ($T_{AddOps} \cdot T_{AddCCs}$) captures the partial sum accumulation cost.

$$T_{Ops} = \begin{bmatrix} N_{FT(C)} + (4 \cdot \text{Shifts} \cdot N_{DT} \cdot N_{FT(C)}) \\ + K + (T_{AddOps} \cdot T_{AddCCs}) \end{bmatrix} \cdot N_{FT(R)} \quad (11)$$

**Equation (12)** defines the peak compute throughput in GFLOPs/sec, capturing the arithmetic intensity under the proposed mapping strategy. The term $(I + 2P/S)^2$ estimates output activations based on the input size $I$, padding $P$, and stride $S$. This is scaled by the number of filters ($N_F$), input channels ($D$), and filter size ($F$ X $F$). The denominator $T_{Ops}$ reflects total execution latency, and the result is scaled by the MAVeC clock frequency ($f_{MAVeC}$) in GHz.

$$\frac{\text{GFLOPs}}{\text{Second}} = \frac{2 \cdot (I + 2P/S)^2 * (N_F \cdot D \cdot F^2)}{T_{Ops}} * f_{MAVeC} \quad (12)$$

System throughput, measured in Kilo Inferences Per Second (**KIPS**), characterizes the end-to-end inference rate achievable on the MAVeC System-on-Chip (SoC) while executing the 7-D convolution workload under the proposed mapping strategy. As shown in **equation (13)**, KIPS is computed by dividing the number of operations executed per second (**Ops/Sec**) by the number of operations required per inference (**Ops/Inf**). The per-inference operation count, given in **equation (14)**, is derived from the total number of operations normalized by batch size (**B**) and the number of images per batch (**N**). **Equation (15)** models the operations-per-second by incorporating the number of available tiles (each comprising 256 PEs), the average PE utilization ($Util_{avg}$), and the operating frequency $f_{MAVeC}$ (in GHz). The total time ($T_{Total} = T_{PCIe} + T_{WL} + T_{MT} + T_{OP}$) is partitioned into four components: PCIe transfer latency from host to off-chip memory ($T_{PCIe}$), weight loading cycles ($T_{WL}$), message transfers cycles ($T_{MT}$), and execution cycles ($T_{OP}$).

$$\text{Throughput} = \frac{\text{Ops/Sec}}{\text{Ops/Inf} \times 10^3} \quad (13)$$

$$\text{Ops/Inf} = \frac{\text{Total Operations}}{B \times N} \quad (14)$$

$$\text{Ops/Sec} = \left(\frac{\text{Ops}_{Total}}{T_{Total}}\right) \times (\text{Tiles} \times 256) \times \frac{Util_{avg}}{100} \times f_{MAVeC} \quad (15)$$

*C. Performance Analysis*

*Figure 7* illustrates the performance trends of the proposed mapping strategy under varying hardware configurations (16X16, 32X32, and 64X64) while processing a set of convolution workloads with increasing complexity. Each convolution operates over a fixed input tensor of size 56X56XD, where both the input depth (**D**) and number of filters (**N_F**) vary over {64, 128, 256, 512}, and the stride (**S**) and padding (**P**) are fixed at **1**. The folding strategy decomposes the filter tensor across the PE array in a weight-stationary fashion, resulting in a collection of partial filter folds per workload, with all spatial output locations processed using an image block, where each image block consists of **P** image folds and shifts along the **Q** dimension.

As filter dimensions scale, the number of required filter folds increases, particularly for smaller PE arrays due to limited spatial capacity per fold, as shown in *Table 3*. For example, the 3X3X512X512 workload yields 16,384 folds on a 16X16 array versus only 824 on a 64X64 array. Despite this disparity, average utilization remains flat at 75% for the 16X16 and 32X32 configurations but climbs sharply to over 92% on the 64X64 array (*Figure 7(a)*). This improvement stems from the ability to spatially map larger and more contiguous fold regions that better align with the tile geometry, reducing fragmentation and idle PEs—even though fold shapes vary across instances.

Table 3. Mapping configuration summary for synthetic convolution workload.

| Workload | PE Array | Mapping Constructs | | | |
|---|---|---|---|---|---|
| | | Filter Fold | | Image Block | |
| | | Count | Type | Length | Shifts |
| 3X3X64X64 | 16 X 16 | 256 | Partial | | |
| 3X3X512X128 | | 1024 | Partial | | |
| 3X3X512X512 | | 4096 | Partial | | |
| 3X3X512X512 | | 16384 | Partial | | |
| 3X3X64X64 | 32 X 32 | 64 | Partial | 56 | 56 |
| 3X3X512X128 | | 256 | Partial | | |
| 3X3X512X512 | | 1024 | Partial | | |
| 3X3X512X512 | | 4096 | Partial | | |
| 3X3X64X64 | 64 X 64 | 13 | Partial | | |
| 3X3X512X128 | | 52 | Partial | | |
| 3X3X512X512 | | 208 | Partial | | |
| 3X3X512X512 | | 824 | Partial | | |

Execution time, shown in *Figure 7(b)* (log scale), accounts for fold traversal, spatial shifts, and accumulation overheads. As array size increases, more operations are processed in parallel, reducing sequential depth-wise computations. For the largest filter, execution time drops from 20.1 million cycles on the 16X16 array to just over 10 million on the 64X64 array, achieving nearly a 2x speedup. Peak compute-throughput, shown in *Figure 7(c)* (log scale) increases with array size due to reduced execution time. Larger arrays support greater parallelism and more efficient fold packing, lowering latency, although the operation count remains fixed across configurations. This leads to throughput gains from ~78 GFLOPs/sec (16X16) to ~1.56 TFLOPs/sec (64X64).

The reuse properties of the proposed dataflow—formulated in Section *IV*, equations (6) to (9)—are empirically validated in *Figure 8*. Temporal reuse of weights, shown in *Figure 8(a)*, grows significantly with array size due to the ability to hold and reuse larger filter folds across all shift cycles within an image block. Spatial reuse of input activations, illustrated in *Figure 8(b)*, benefits from increased horizontal multicasting and longer reuse windows enabled by deeper PE arrays. *Figure 8(c)* quantifies the accumulated partial sums through spatial reduction, which scales with both the number of active folds and the parallel compute lanes. These trends confirm that larger arrays enhance reuse and aggregation efficiency by better aligning with fold geometry, thereby reducing data movement and improving throughput under the proposed dataflow.

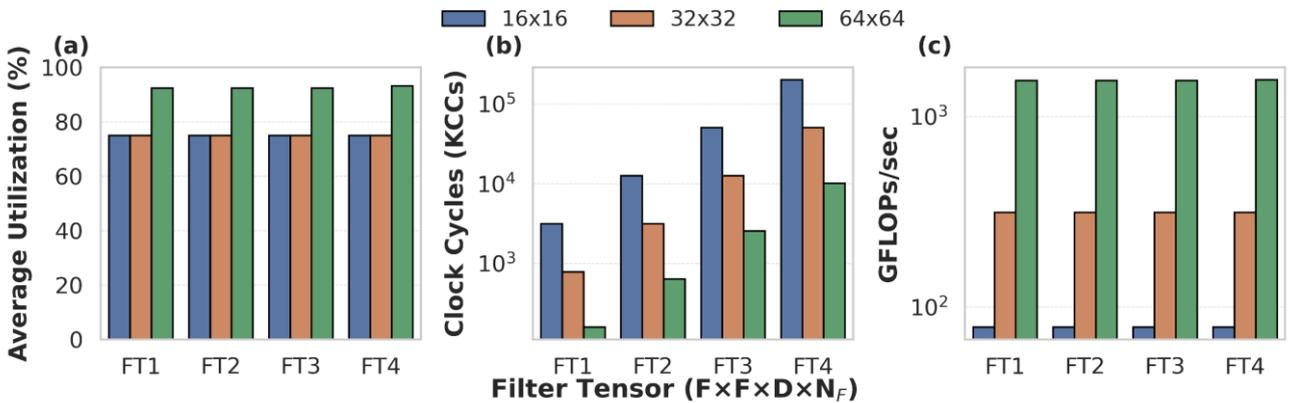

**Figure 7.** Performance trends of the proposed mapping strategy under varying hardware configurations. **(a)** Average Utilization improves with PE array due to better fold alignment and reduced fragmentation. **(b)** Execution time decreases with larger arrays due to parallelism. **(c)** Compute throughput increases significantly with array size due to faster fold processing.

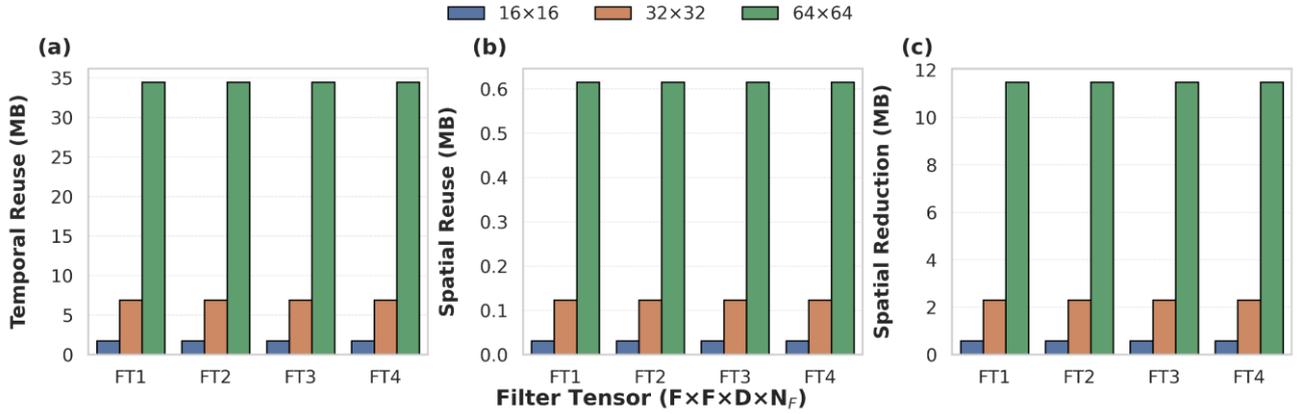

**Figure 8. Data reuse trends under the proposed mapping strategy across hardware configurations. (a)** Temporal reuse increases with array size as more hardware rows accommodate additional filters per fold **(b)** Spatial reuse improves due to broader multicasting of input activations across wider PE arrays, enhancing reuse across shift cycles. **(c)** Spatial reduction improves as broader arrays reduce more columns and depth slices concurrently.

*Figure 9* presents a layer-wise breakdown of average utilization and total clock cycles across the 13 convolutional layers of VGG-16 for three PE array configurations. In *Figure 9(a)*, the 64X64 array consistently achieves over 90% utilization across layers by accommodating wider, denser folds with minimal fragmentation. In contrast, 16X16 and 32X32 arrays remain capped near 75%, with a noticeable drop in early layers like conv1_2. As shown in *Figure 9(b)*, this utilization gap translates to lower execution cycles per layer on the larger array, driven by reduced fold counts and greater parallelism across active PEs.

The system throughput for VGG-16 is evaluated under the proposed mapping strategy using a 64X64 PE array operating at 1 GHz, with GDDR7 off-chip memory (4.5 GB/s) and a PCIe Gen6 x16 host interface (126 GB/s). This setup achieves 12.7 KIPS, reflecting effective coordination of operation and communication stages. Total runtime includes 7.6 million cycles for PCIe transfer, 0.64 million for weight loading, 260.7 million for message movement, and 21.1 million for compute execution. Despite the dominance of message transfer, the system sustains high throughput by enabling parallel fold execution, maximizing on-chip spatial reuse, minimizing off-chip access, and effectively exploiting data locality through compact fold placement.

## VI. CONCLUSION

This work presents a structured mapping framework that translates the canonical seven-dimensional convolution loop nest into a set of hardware-aware execution primitives. By decomposing the convolutional workload into Filter Folds, Image Blocks, and Image Folds, the proposed strategy exposes parallelism, reuse, and reduction opportunities explicitly aligned with spatial dataflow architectures. Implemented on the MAVeC accelerator, this mapping leverages weight-stationary execution, spatial multicasting, and hierarchical partial sum accumulation to optimize data movement and maximize utilization. Experimental results on VGG-16 and synthetic benchmarks demonstrate high PE utilization, reduced execution latency, and scalable throughput across varying array sizes. The methodology offers a principled approach to aligning loop-level semantics with spatial hardware execution, enabling efficient

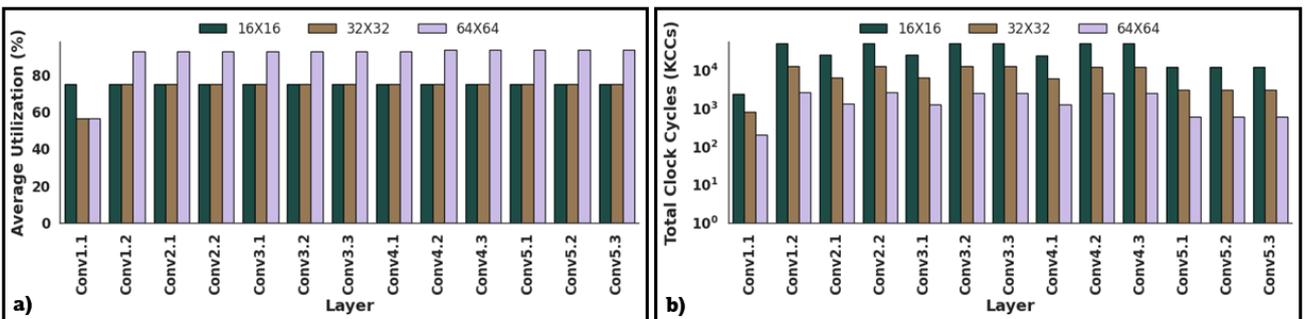

**Figure 9. Layer-wise performance breakdown for VGG-16 across different PE array sizes. (a)** Average utilization improves with larger arrays, where 64×64 consistently exceeds 90% due to better fold packing and minimal fragmentation. **(b)** Total clock cycles decrease layer by layer on larger arrays, benefiting from reduced fold counts and increased parallelism.

deployment of deep learning workloads on programmable accelerators.